\documentclass[intlimits,twoside,a4paper]{article}

\usepackage{amsmath,amssymb}
\usepackage{graphicx}
\usepackage{wrapfig}

\usepackage[T2A]{fontenc}
\usepackage[cp1251]{inputenc}

\usepackage[eqsecnum]{cmpj}

\issue{2011}{14}{2}{23702}

\doinumber{10.5488/CMP.14.23702}

\title[Spectral densities and diagrams of states of 1D ionic Pauli conductor]
{Spectral densities and diagrams of states of one-dimensional ionic Pauli conductor}

\author[I.V.~Stasyuk, O.~Vorobyov, R.Ya.~Stetsiv]{I.V.~Stasyuk, O.~Vorobyov, R.Ya.~Stetsiv}

\address{Institute for Condensed Matter Physics of the National
Academy of Sciences of Ukraine, 1~Sventsitskii Str., 79011~Lviv,
Ukraine}

\authorcopyright{I.V.~Stasyuk, O.~Vorobyov, R.Ya.~Stetsiv, 2011}
\date{Received January 12, 2011, in final form February 14, 2011}

\begin{document}

\maketitle

\begin{abstract}
We focus on the features of spectra and diagrams of states
obtained via exact diagonalization technique for finite ionic
conductor chain in periodic  boundary conditions. One dimensional
ionic conductor is  described with the lattice model where ions
are treated  within the framework of ``mixed'' Pauli statistics.
The ion transfer and nearest-neighbour interaction between ions
are taken into account. The spectral densities and diagrams of
states for various temperatures and values of interaction are
obtained. The conditions of transition from uniform (Mott
insulator) to the modulated (charge density wave state)  through
the superfluid-like state (similar to the state with the
Bose-Einstein condensation observed in hard-core boson models) are
analyzed.

\pacs 75.10.Pq, 66.30.Dn, 66.10.Ed
\end{abstract}

\section{Introduction}

Ionic conductors are a wide class of physical and biological
objects ranging from ice to DNA membranes. One of the most
interesting subclasses of these are superionic conductors that
exhibit high temperature phase with high conductivity that arises
due to the motion of ions~\cite{tomoyose} or
protons~\cite{belushkin}. Theoretical description of systems with
ionic conductivity is most frequently based on the lattice models.
Some of them treat ions as Fermi-particles focusing on different
aspects of the ionic subsystem like long-range
interactions~\cite{salejda1,salejda2,pavlenko1} or interaction
with phonons~\cite{pavlenko2,tomchuk}. Some recent attempts have
also been made towards short-range interactions between
particles~\cite{our2,our3,our4,our5,our5a}.

However, a more correct consideration of ions should be based on
the mixed statistics of Pauli~\cite{mahan} since these particles
are bosons by nature but they also obey the Fermi rule. Due to the
special commutation rules, the utilization of Pauli operators
generates additional mathematical complexities. On the other hand,
this approach might be very effective. For instance, it has been
shown that the lattice model of Pauli particles is capable of
describing the appearance of superfluid-like state (that
corresponds to superionic phase) in the system even in the absence
of interaction between particles~\cite{dulepa,dulepa2,micnas}. On
the other hand, the lattice model of Pauli particles is similar to
the hardcore Bose-Hubbard model widely used for the description of
ionic conductivity phenomena as well as for the modelling of
energy spectrum of absorbed  ions on a crystal surface and
intercalation in crystals~\cite{mysakovych}. Bose-Hubbard model
also exhibits the transition from Mott insulator state to
superfluid-like
state~\cite{mysakovych2,batrouni1,micnas2,batrouni2,batrouni3,batrouni4,schmid}.
Some of the authors also observe the possibility of formation of
intermediate ``supersolid'' phase that may appear on the phase
diagrams alongside the transition from dielectric (CDW) to
superfluid phase.

In this work we focus on the diagrams of state for one-dimensional
ionic conductor described by the system of Pauli particles. Our
lattice model includes ion transfer as well as the interaction
between nearest-neighbouring ions. We calculate the
single-particle spectral densities of the finite system in
periodic boundary conditions and obtain the diagrams of state
analyzing the features of this spectra. The conditions of
transition from Mott insulator (MI)-like state to the
 modulated charge density wave (CDW) state through the  superfluid(SF)-like state
 (similar to the state with the Bose-Einstein (BE) condensation  observed in
 hard-core boson models) are discussed.

\section{The model for ionic conductor}

Let us consider the chain of heavy immobile ionic groups (large
circles in figure~1) and light ions that move along this chain
occupying positions denoted by small circles in figure~1. The
subsystem of light ions is described with the following
Hamiltonian
\begin{eqnarray}
\hat{H} &=& t \sum\limits_{i} (c_{i}^{+}c_{i+1} + c_{i+1}^{+}
c_{i}) + V \sum_{i}n_{i}n_{i+1} - \mu \sum_{i} n_i\, \label{ham1}.
\end{eqnarray}
This model takes into account the nearest-neighbour ion transfer
(with hopping parameter $t$) and interaction between ions that
occupy nearest-neighbouring positions (with corresponding
parameter $V$). If this Hamiltonian is considered within the
framework of Fermi statistics, the corresponding model is known as
spinless-fermion  model. This model is widely used in the theory
of strongly correlated electron systems~\cite{vlaming} as well as
for   the description of ionic conductors~\cite{lorenz}. A more
complex two-sublattice case of this model can be applied to a
proton conductor~\cite{or-tun}. More correct consideration of ions
should be based on ``mixed'' Pauli statistics and this approach is
used onwards. In this case, the model~(\ref{ham1}) is equivalent
to the extended hard-core boson model, i.e., boson Hubbard model
with repulsive interaction between nearest neighbours and infinite
on-site repulsion~\cite{batrouni}. The latter is often applied to
the investigation of the problems of BE-condensation and
superfluidity.
\begin{figure}[h]
\centerline{\includegraphics[width=0.95\columnwidth]{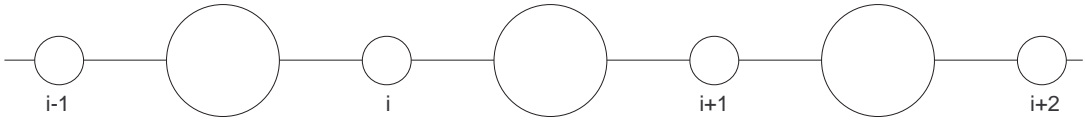}}
\caption{The model for one-dimensional ionic conductor. Large
circles denote heavy ionic groups while the small ones denote
light movable ions.} \label{model}
\end{figure}

\section{Exact diagonalization technique}

We calculate the spectral densities of one-dimensional ionic Pauli
conductor using exact diagonalization technique. For the chain of
$N$ sites, we introduce the many-particle states
\begin{equation}
\mid n_{1,a} n_{1,b} \ldots n_{N,a} n_{N,b} \rangle .
\end{equation}
The Hamiltonian matrix on the basis of these states is the matrix
of the order $2^N \times 2^N$ and is constructed  as follows:
\begin{eqnarray}
H_{mn} = \sum \limits_{i=1}^N \left[ t \left(
H^{(1)}_{mn} + H^{(2)}_{mn} \right) + \widetilde{V} H^{(3)}_{mn} - \mu H^{(4)}_{mn}
\right]  , \label{matrixel}
\end{eqnarray}
where
\begin{eqnarray}
H^{(1)}_{mn} &=& \langle n_{1} \ldots | c^+_{i} c_{i+1} |
n'_{1} \ldots \rangle = \delta(n_{i} \!-\! n'_{i} \!-\! 1)
\delta(n_{i+1} \!-\! n'_{i+1} \!+\! 1) \nonumber \\ && \times \prod\limits_{l \neq i ;
i+1} \delta(n_l \!-\! n'_l), \nonumber
\\ H^{(2)}_{mn} &=& \langle n_{1} \ldots | c^+_{i+1} c_{i} |
n'_{1} \ldots \rangle = \delta(n_{i} \!-\! n'_{i} \!+\! 1)
\delta(n_{i+1} \!-\! n'_{i+1} \!-\! 1) \nonumber \\ && \times \prod\limits_{l \neq i ;
i+1} \delta(n_l \!-\! n'_l), \nonumber \\
H^{(3)}_{mn} &=& \langle n_{1} \ldots | n_{i} n_{i+1} | n'_{1}
\ldots \rangle = \delta(n_{i} \!-\! 1) \delta(n'_{i} \!-\! 1)
\delta(n_{i+1} \!-\! 1)  \nonumber \\ && \times \delta(n'_{i+1} \!-\! 1) \prod\limits_{l
\neq i ; i+1} \delta(n_l \!-\! n'_l), \nonumber \\
H^{(4)}_{mn} &=& \langle n_{1} \ldots | n_{i} | n'_{1}
\ldots \rangle =
 \delta(n_{i} \!-\! 1) \delta(n'_{i} \!-\! 1)
\prod\limits_{l \neq i} \delta(n_l \!-\! n'_l). \nonumber
\end{eqnarray}

This matrix is diagonalized numerically
\begin{eqnarray}
U^{-1} H U = \widetilde{H} = \sum \limits_p \lambda_p
\widetilde{X}^{pp},
\end{eqnarray}
where $\lambda_p$ are eigenvalues of the Hamiltonian, $\widetilde{X}^{pp}$ are Hubbard-operators. The same transformation is applied to the creation and annihilation operators
\begin{eqnarray}
U^{-1} c_{i} U = \sum \limits_{pq} A_{pq} \widetilde{X}^{pq} \, ,
\qquad U^{-1} c^+_{i} U = \sum \limits_{pq} A^*_{rs}
\widetilde{X}^{rs}
\end{eqnarray}
which are required to construct one-particle Green's function $\ll
c_{i,a} | c^+_{i,a} \gg$ that contains information about
one-particle energy spectrum of the system. For Pauli creation and
annihilation operators, this Green's function can be constructed
in two ways, i.e., commutator Green's function
\begin{eqnarray}
\ll c_i (t) | c^+_i (t') \gg = - {\rm i} \Theta (t-t') \langle
[c_i (t), c^+_i (t')] \rangle \label{grcom}
\end{eqnarray}
and anticommutator Green's function
\begin{eqnarray}
\ll c_i (t) | c^+_i (t') \gg = - {\rm i} \Theta (t-t') \langle \{
c_i (t), c^+_i (t') \} \rangle. \label{grant}
\end{eqnarray}
Imaginary part of these Green's functions are one-particle spectral
densities (also referred to as densities of states or DOS)
\begin{eqnarray}
\rho(\omega) &=& - \frac{1}{\pi N} \sum \limits_{i=1}^N
\textrm{Im} \ll c_{i,a} | c^+_{i,a} \gg \nonumber \\ &=& -
\frac{1}{\pi N} \sum \limits_{i=1}^N \textrm{Im} \left[\frac{1}{Z}
\sum \limits_{pq} A_{pq} A^*_{pq} \frac{{\rm e}^{-\beta \lambda_p}
- \eta {\rm e}^{-\beta \lambda_q}}{\omega -
(\lambda_{q}-\lambda_{p})} \right], \label{greenf}
\end{eqnarray}
where $Z = \sum_p {\rm e}^{-\beta \lambda_p}$. Spectral densities
in (\ref{greenf}), obtained from commutator $\eta = 1$
(\ref{grcom}) and anticommutator $\eta = -1$ (\ref{grant}) Green's
functions, respectively, exhibit a discrete structure, i.e.,
consist of several $\delta$-peaks due to the finite size of a
cluster. Therefore, we apply the periodic boundary conditions to
the cluster and introduce a small parameter $\Delta$ to broaden
the $\delta$-peaks according to Lorentz distribution
\begin{equation}
\delta (\omega) \rightarrow \frac{1}{\pi}
\frac{\Delta}{\omega^2+\Delta^2}\,.
\end{equation}

\vspace{5mm}

\section{Results and discussion}

\begin{figure}[h]
\centerline{\includegraphics[width=0.48\columnwidth]{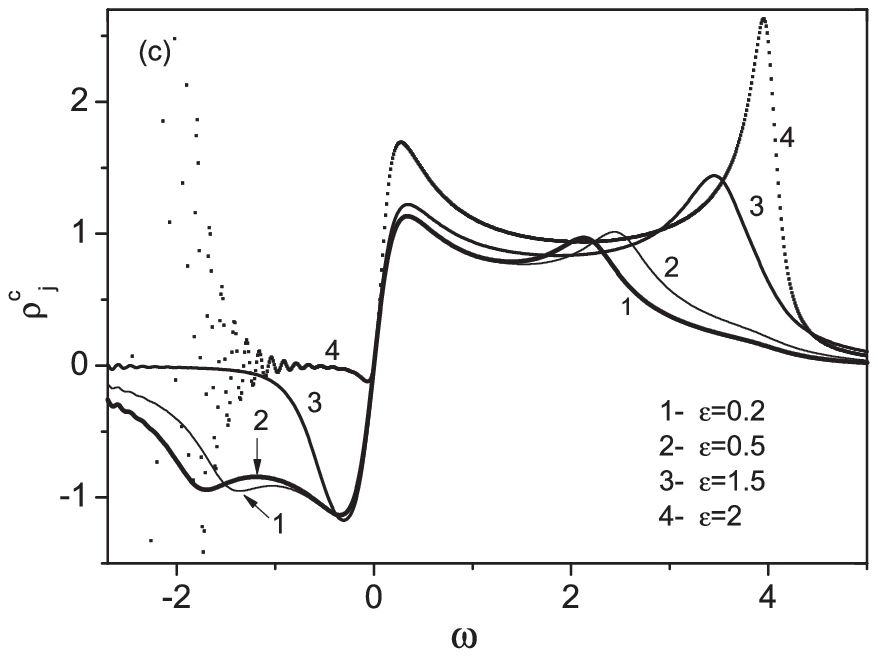}
\includegraphics[width=0.46\columnwidth]{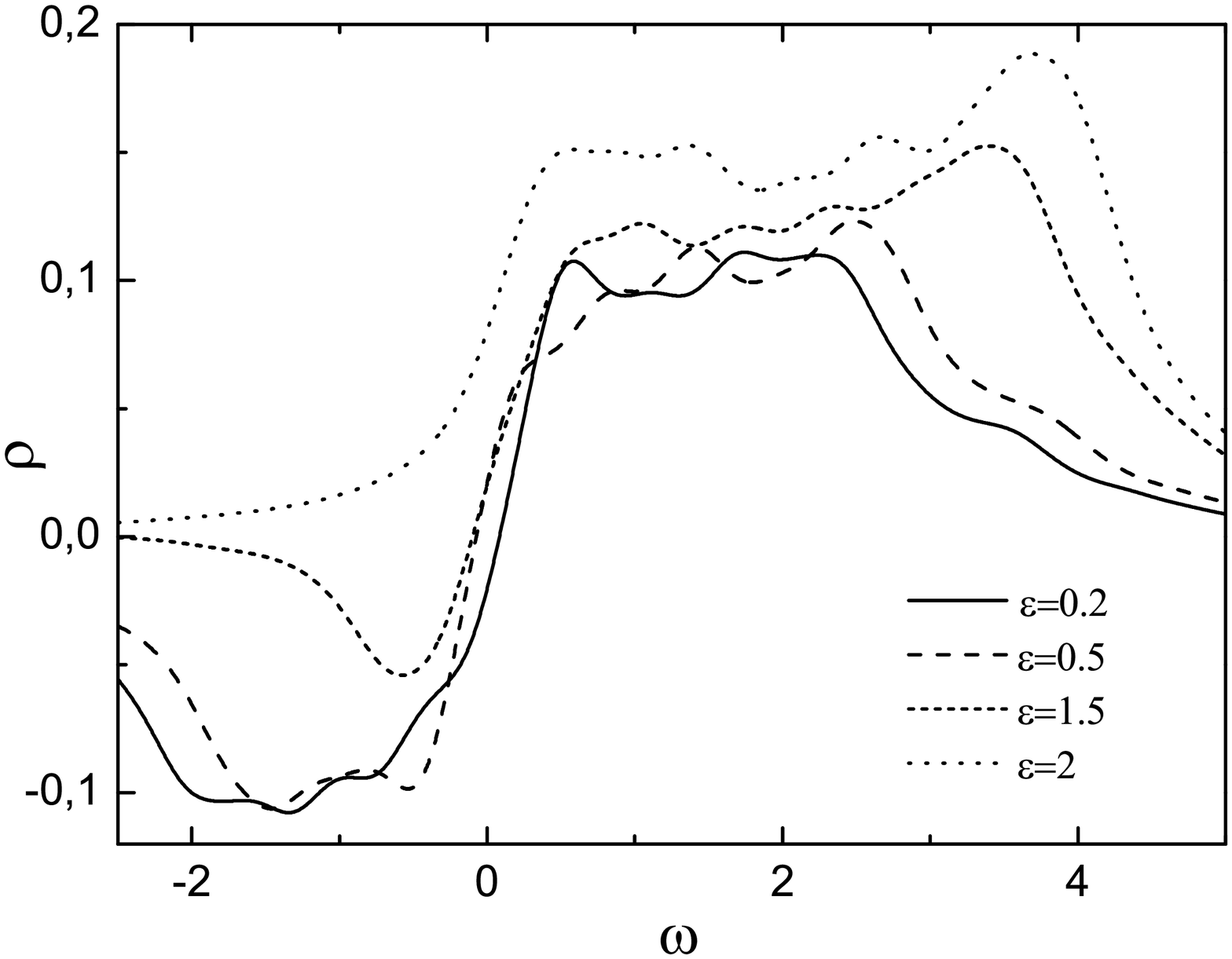}}
\caption{Commutator spectral
densities of non-interacting ($V =0$) ionic Pauli conductor for different values of
chemical potential (right figure, $\varepsilon=-\mu$) compared to exact results obtained in~\cite{dulepa} via fermionization
procedure (left figure, $\varepsilon=\varepsilon_{0}-\mu$). $t = 1, T=0.2, \Delta = 0.4$. Spectral density on the left figure is scaled to 2$\pi$,
while on the right figure it is scaled to unity.} \label{dos-mu}
\end{figure}

\begin{figure}[h]
\vspace{10mm}
\centerline{\includegraphics[width=0.5\columnwidth]{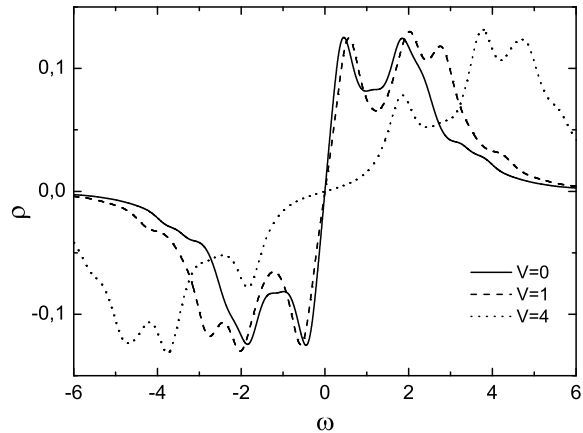}}
\caption{Commutator spectral density of ionic Pauli conductor for
different values of nearest-neighbour interaction $V$ at half
filling ($\mu=0$). $t = 1, \, T=0.2, \, \Delta = 0.4$.}
\label{dos-V}
\end{figure}

We perform calculations of one-particle spectral densities of
one-dimensional ionic Pauli conductor (\ref{ham1}) for the chain
of ten sites ($N=10$) in periodic boundary conditions. To test the
results we compare them with the exact solution obtained by means
of fermionization procedure~\cite{dulepa} in the absence of
nearest neighbour interaction ($V =0$) for different values of
chemical potential (figure~\ref{dos-mu}). The redistribution of
statistical weight with the change of chemical potential level can
be observed, and a good level of agreement is achieved. The
analysis of spectral densities is a way to distinguish the
different states of an ionic subsystem and the corresponding
conditions. In the case of half-filling, as we turn on the
interaction starting with spectral density whose shape corresponds
to SF state, we observe the development of the gap in the spectra
(figure~\ref{dos-V}). A similar effect was found for ionic and
proton conductors described by the similar models within the
framework of Fermi statistics. It was connected with the splitting
of spectra due to charge ordering with doubling of the lattice
period~\cite{our5,our5a}. The detailed analysis of spectral
densities of Fermi and Pauli models of ionic conductor can be
found in~\cite{vorobyov}. So, here we have a transition from SF to
CDW state. This transition was described analytically for $T=0$
in~\cite{kuhner,hen}. We have investigated the case of $T \neq 0$
when this transition manifests itself as a crossover.

\begin{figure}[!h]
\centerline{\includegraphics[width=0.48\columnwidth]{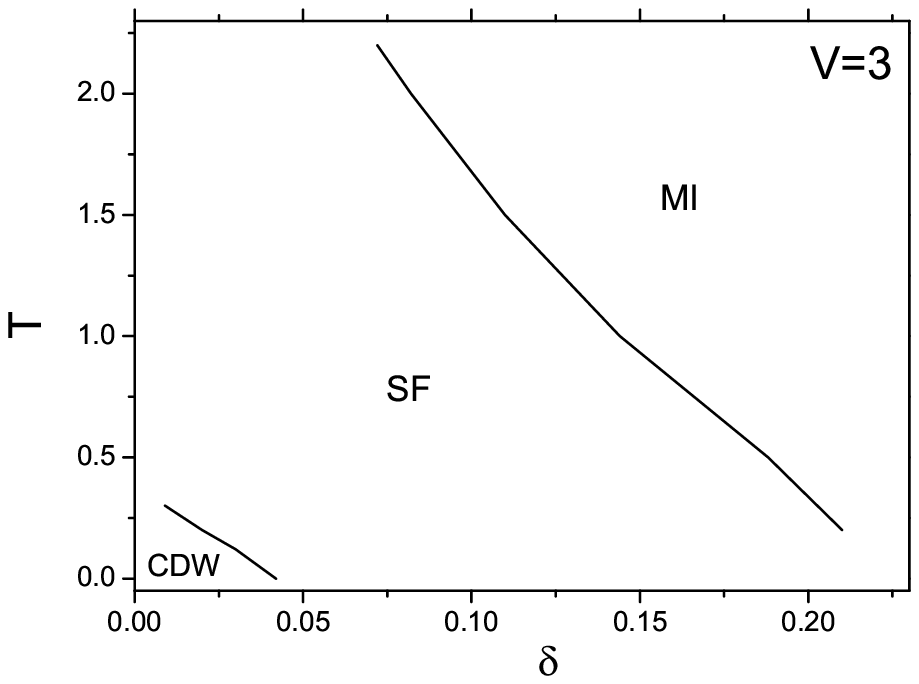}
\includegraphics[width=0.48\columnwidth]{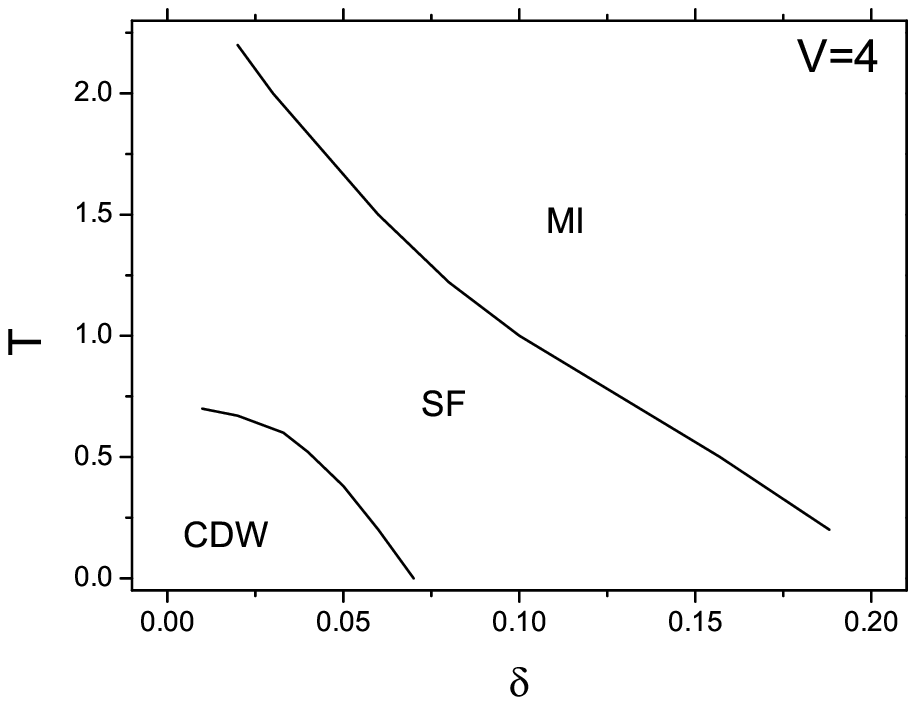}}
\vspace{5mm}
\centerline{\includegraphics[width=0.49\columnwidth]{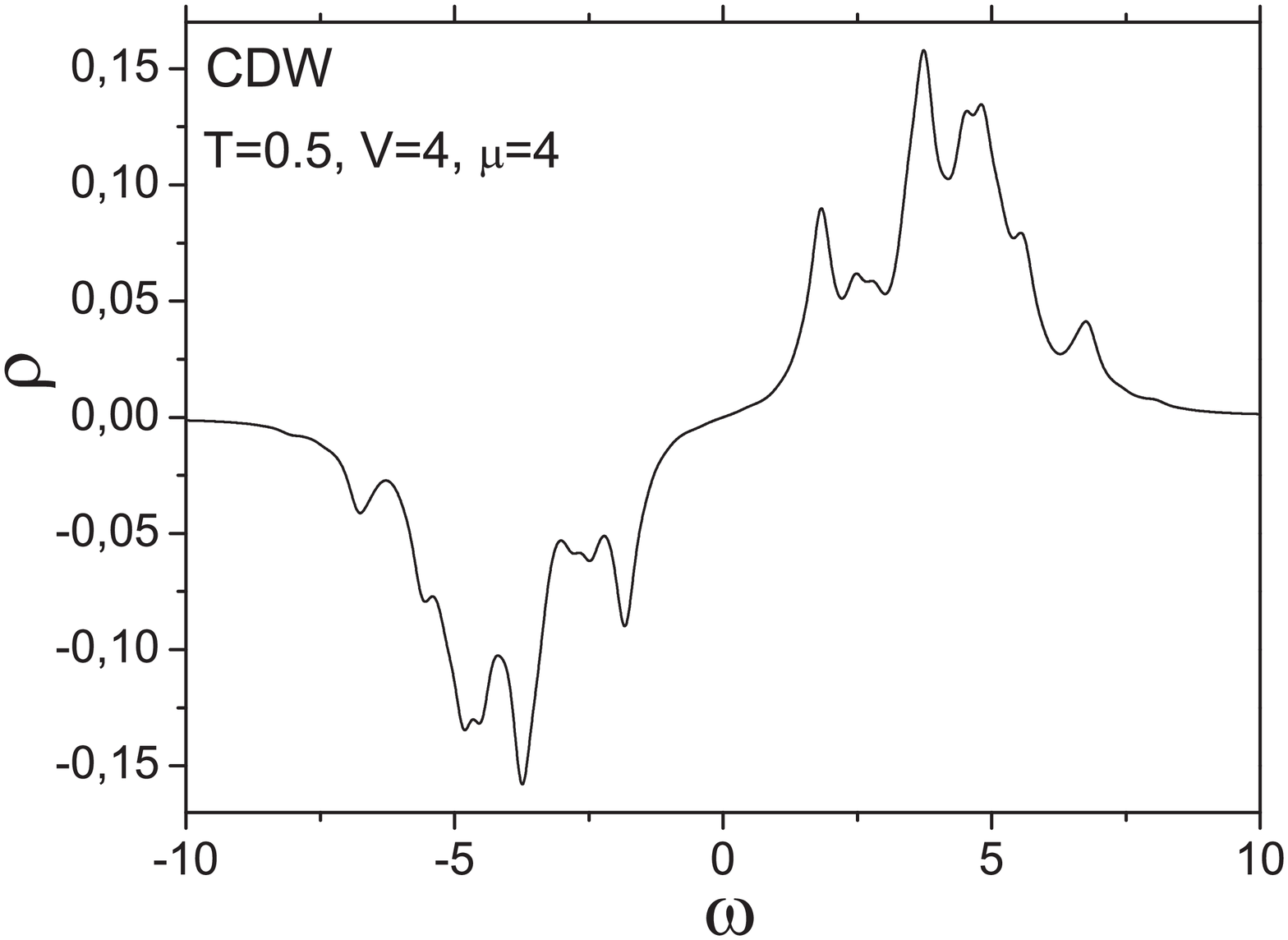}   
\hspace{-0.02\columnwidth}
\includegraphics[width=0.48\columnwidth]{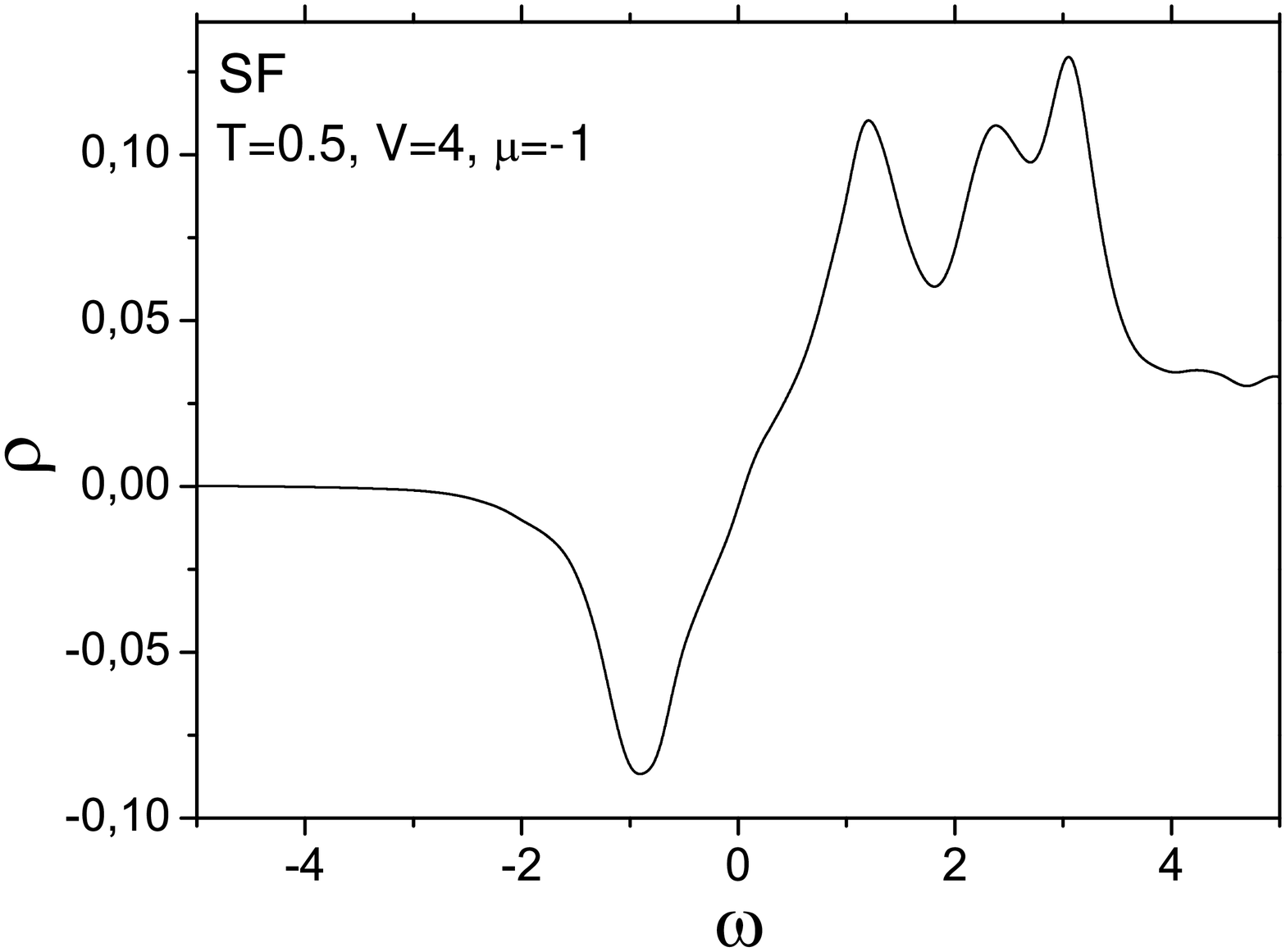}}              
\vspace{5mm}
\centerline{\includegraphics[width=0.5\columnwidth]{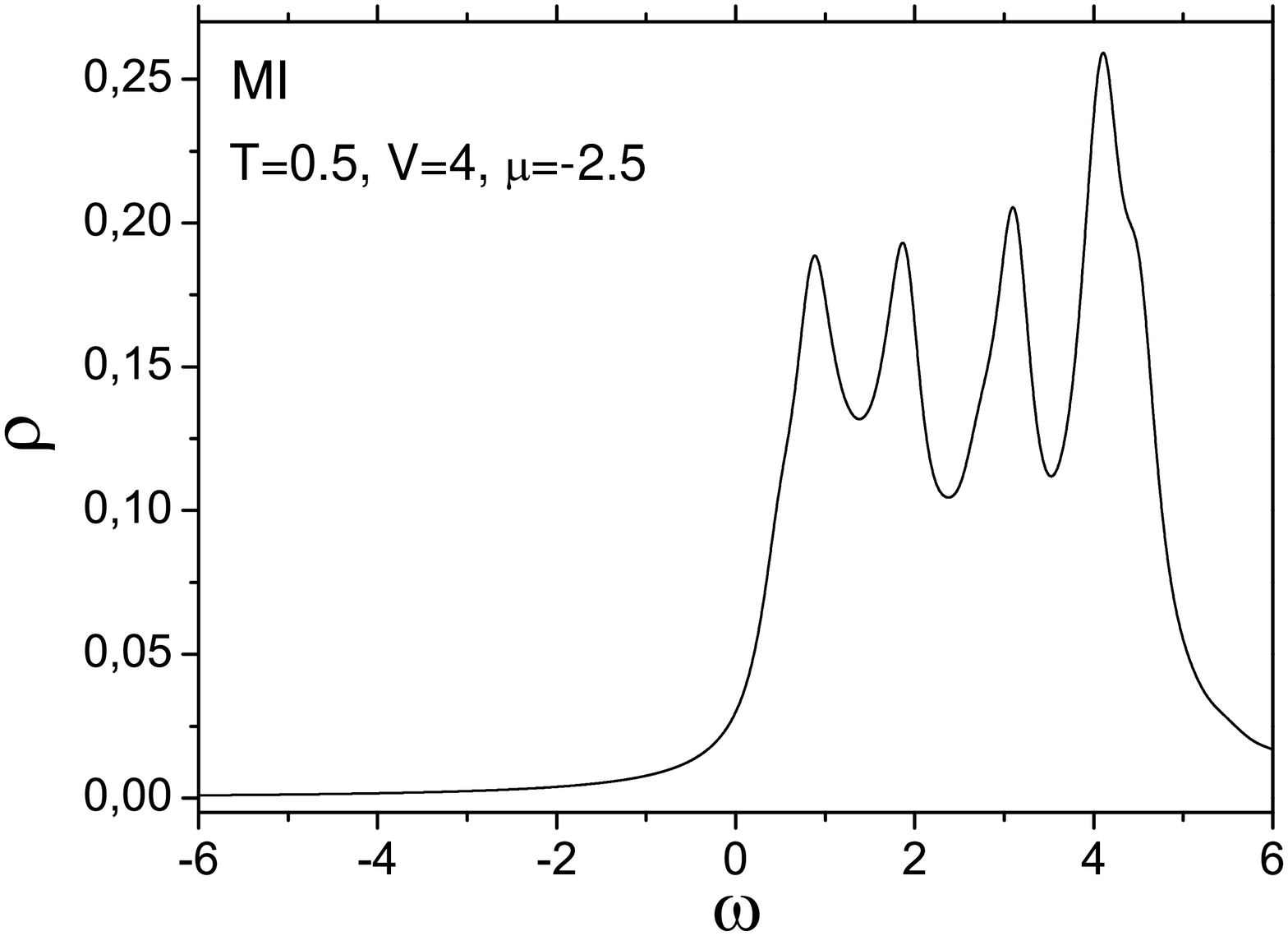}}  
\caption{Diagrams of state for different interactions and the spectral densities
for negative $\delta$ that correspond to CDW, SF and MI states. $t = 1, \Delta = 0.25$.
$\delta=\langle n\rangle-1/2$ denotes the deviation from half-filling.}
 \label{diagofs}
\end{figure}

\begin{figure}
\centerline{\includegraphics[width=0.46\columnwidth]{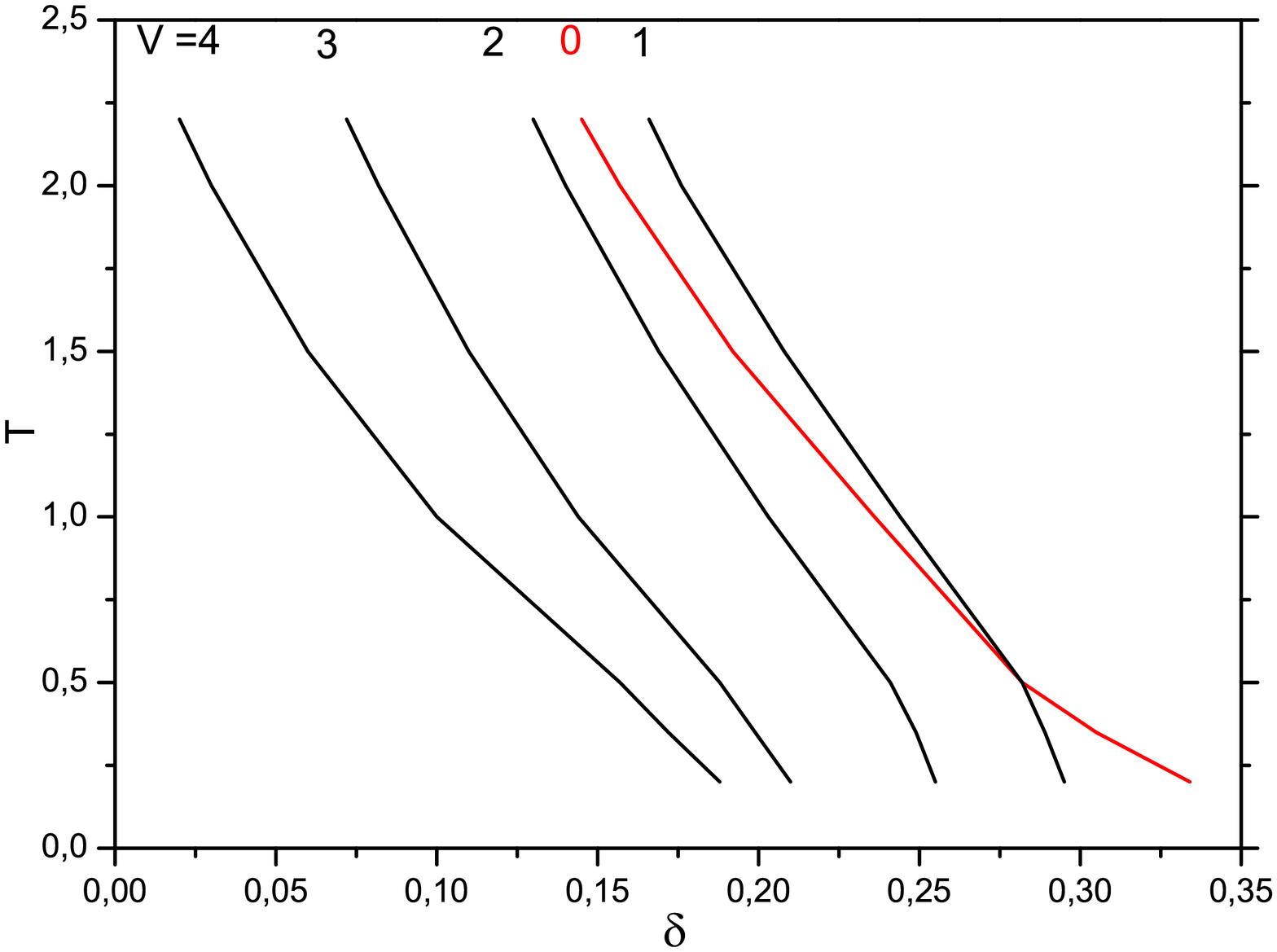}
\includegraphics[width=0.46\columnwidth]{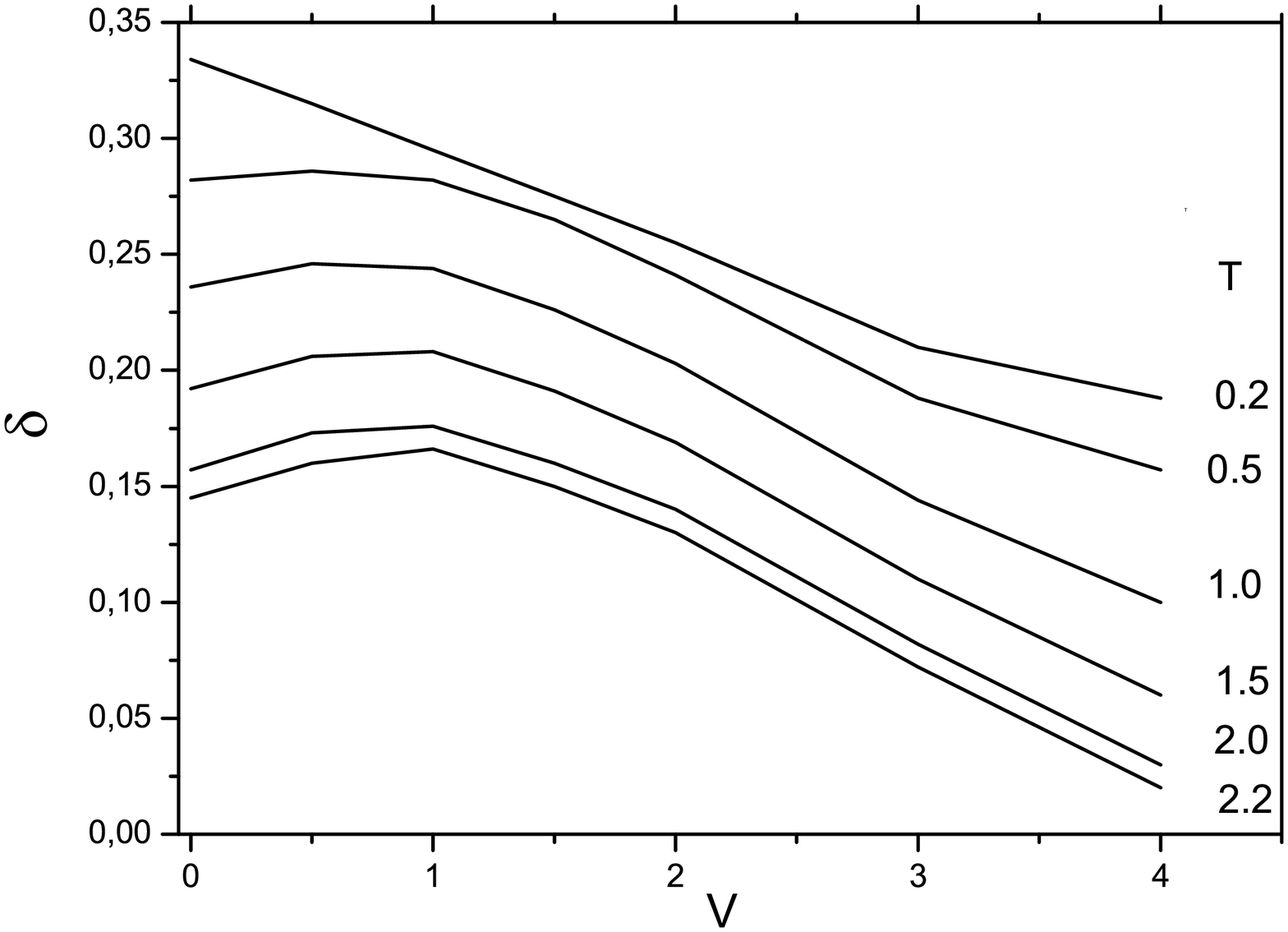}}
\caption{The curves that separate SF- and MI- states at different
temperatures and values of interaction. SF-state is on the left
hand side of the curves and below them. $t = 1,\, \Delta = 0.25$.
$\delta=\langle n\rangle-1/2$ denotes the deviation from half-filling.}
\label{diagsf}
\end{figure}

\looseness=1The existence of a gap in the spectra which separates the bottom
of the energy band from the chemical potential level is also the
sign of the presence of homogenous MI state. The vanishing of the
gap and the appearance of a negative branch points to the
transition to SF-like state (see, for example,~\cite{menotti}).
According to these criteria we analyze the spectral densities at
different temperatures and values of interaction and build the
corresponding diagrams of state (figure~\ref{diagofs}). The system
is in MI homogenous state at high temperatures and far away from
half-filling (at large $\delta$). As the temperature decreases or
one comes closer to half-filled case, the system undergoes
transition to SF-like state which corresponds to the appearance of
the negative branch without any gap on the spectral density. At
further temperature decrease and closer to half filling, we
observe a transition to the state with the gap on the spectral
density that corresponds to the CDW-ordering (the negative branch
still exists). As the interaction strength $V$ increases, such a
region becomes broader while the region of SF-like state becomes
smaller. On the other hand, with the decrease of~$V$, the CDW
state diminishes and disappears at $V \approx t$. It should be
mentioned that the sequence of states that the system is going
through at the increase of mean occupancy $\delta$, corresponds to
the phase diagram obtained in~\cite{kuhner,hen,hen2}. We have also
performed a detailed analysis of the transition to SF-like state
at different temperatures and values of interaction
(figure~\ref{diagsf}). It is interesting that at weak interactions
($V<1$), the increase of interaction strength facilitates the
formation of SF-like state while further increase of $V$
suppresses this transition. The shift of the curves that separate
MI- and SF-states towards smaller values of $\delta$ with the
increase of interaction strength that we have obtained is also
observed on the phase diagrams obtained by other authors
\cite{kuhner,hen}.

\section{Conclusions}

We have performed the analysis of diagrams of states of
one-dimensional Pauli ionic conductor using an exact
diagonalization technique. We have shown that the system undergoes
transition from Mott insulator to superfluid-like state and then
to CDW-sate. At weak interaction, the latter transition may
vanish.


\ukrainianpart
\title{Спектральні густини та діаграми стану одновимірного
іонного провідника Паулі}
\author{І.В. Стасюк, О. Воробйов, Р.Я. Стеців}
\address{Інститут фізики конденсованих систем НАН України, вул.~І.~Свєнціцького, 1, 79011 Львів, Україна}
\makeukrtitle
%
%
%
%
\begin{abstract}
\tolerance=3000%
Робота присвячена вивченню енергетичного спектру та діаграм станів, отриманих методом точ\-ної
діагоналізації для скін\-чен\-ного іонного ланцюгового провідника в періодичних гра\-нич\-них умовах.
Одновимірний іонний провідник описується ґрат\-ко\-вою мо\-дел\-лю, де іони розглядаються як частинки Паулі, при
цьому вра\-хо\-ву\-єть\-ся іонний перенос і двочастинкова взаємодія між най\-ближ\-чи\-ми сусі\-дами. Було
розраховано та проаналізовано спектральні густини та діаграми стану такої системи для різ\-них температур та
величин взаємодії. Про\-аналіз\-ов\-ано умови переходу системи з од\-но\-рід\-но\-го (стану т. зв. моттівського
діелектрика) у модульований стан через стан типу фази з бозе-конденсатом (по\-діб\-ної до надплинної фази в
моделях жорстких бозонів).
\keywords статистика Паулі, густина станів, іонний провідник
\end{abstract}

\end{document}